\documentclass[11pt,a4paper]{article}
\pdfoutput=1
\usepackage{jcappub}

\usepackage{braket}

\usepackage{environ}
\NewEnviron{splitequation}{%
	\begin{equation}\begin{split}
		\BODY
	\end{split}\end{equation}}

\newcommand{\mr}[1]{\mathrm{#1}}

\newcommand{\e}{\mathrm{e}}
\newcommand{\diff}{\mathrm{d}}

\newcommand{\conj}[1]{#1^*}

\newcommand{\cB}{\mathcal{B}}
\newcommand{\cT}{\mathcal{T}}

\newcommand{\cO}{\mathcal{O}}

\newcommand{\cA}{\mathcal{A}}
\newcommand{\cH}{\mathcal{H}}

\providecommand{\abs}[1]{\lvert#1\rvert}

\newcommand{\ddelta}{\delta_\mr{D}}

\newcommand{\Pp}{\mathcal{P}_{\Phi}}

\newcommand{\Bp}{\mathcal{B}_{\Phi}}

\title{Weak lensing corrections to tSZ-lensing cross correlation}
\author{Tilman Tr\"oster,}
\author{Ludovic Van Waerbeke}
\affiliation{Department of Physics and Astronomy, The University of British Columbia, 6224 Agricultural Road, Vancouver, B.C., V6T 1Z1, Canada}
\emailAdd{troester@phas.ubc.ca}
\emailAdd{waerbeke@phas.ubc.ca}

\arxivnumber{}

\abstract{The cross correlation between the thermal Sunyaev-Zeldovich (tSZ) effect and gravitational lensing in wide field has recently been measured. It can be used to probe the distribution of the diffuse gas in large scale structure, as well as inform us about the missing baryons. As for any lensing-based quantity, higher order lensing effects can potentially affect the signal. Here, we extend previous higher order lensing calculations to the case of tSZ-lensing cross correlations. We derive terms analogous to corrections due to the Born approximation, lens-lens coupling, and reduced shear up to order $\cO(\Phi^4)$ in the Newtonian potential. Redshift distortions and vector modes are shown to be negligible at this order. We find that the dominant correction due to the reduced shear exceeds percent-level only at multipoles of $\ell \gtrsim 3000$.}

\begin{document}

\maketitle

\section{Introduction}
Even though both the direct detection of dark matter and its microscopic description have proven to be elusive so far, its macroscopic behaviour is thought to be well understood. The large scale clustering of dark matter has been observed though gravitational lensing and has been found to agree with theoretical predictions, see \cite{Bartelmann2010} for a review and \cite{Hildebrandt2014} for an overview of the recent results with the Canada France Hawaii Telescope Lensing Survey (CFHTLenS). On small scales its clustering behaviour has been modelled with N-body simulations to relatively high precision \cite{Harnois-Deraps2014a}. Conversely, even though the microscopic behaviour of baryons is fully understood, half of the universe's baryon content is in a hitherto unobserved state \cite{Fukugita2004,Bregman2007}; the missing baryon problem.

A significant fraction of these missing baryons might reside in a warm, low density phase beyond galactic halos\cite{Anderson2010}. By cross correlating thermal Sunyaev-Zeldovich (tSZ) effect maps, which traces warm electrons, and mass maps derived from weak gravitational lensing data, there is now observational support for the possibility that a significant fraction of the baryons indeed reside in such a phase \cite{VanWaerbeke2014,Ma2014,Hill2014}.

Future surveys with large sky coverage \cite{DeJong2013,Sanchez2010} will produce data whose precision warrants a more sophisticated theoretical treatment than has been necessary so far. In this work we investigate the effect of higher order lensing terms on the tSZ-lensing cross correlation. There has been considerable effort to characterize higher order contributions to correlations of lensing observables \cite{Schneider1998, Cooray2002, Hirata2003a, Dodelson2005, Dodelson2006, Shapiro2006, Shapiro2009, Krause2010, Bernardeau2010, Vanderveld2011, Bernardeau2012, Andrianomena2014}. Some of these higher order effects, like the rotation power spectrum, have been successfully observed in high-resolution ray-tracing simulations \cite{ Hilbert2009, Becker2013}. Building on this corpus of previous work, we derive analogous contributions to the tSZ-lensing cross correlation.

In section \ref{sec:first-order} we introduce the notation and recapitulate the first order results. In section \ref{sec:correction} we investigate the higher order corrections. The terms related to the Born approximation, i.e., the evaluation of the integrals along the unperturbed photon path, and lens-lens coupling are derived in section \ref{sec:born-lens-lens}. The observed quantity in weak gravitational lensing is the reduced shear. Corrections due to its the non-linear relation to the shear and convergence are derived in section \ref{sec:reduced-shear}. We also consider redshift distortions in section \ref{sec:redshift-distortions} and vector modes in section \ref{sec:vector-modes} and show that they do not contribute to the order we are considering.

\section{First order}
\label{sec:first-order}
In the Newtonian gauge the perturbed Robertson-Walker (RW) metric without anisotropic stresses can be written as
\begin{equation}
	\diff s^2 = a(\eta)^2\left[ -(1+2\Phi)\diff \eta ^2 + (1-2\Phi)\left(\diff \chi^2 + d_A(\chi)^2\diff \Omega^2\right)\right] \ ,
\end{equation}
with $d_A(\chi)$ the comoving angular diameter distance and $\chi$ the comoving radial distance. We will henceforth work in units where $c=1$. The potential $\Phi$ is assumed to be small, i.e., $\Phi << 1$. The first order solution to the geodesic deviation equation at a comoving distance $\chi$ from the observer is then \cite{Schneider1992,Schneider1998,Bartelmann2010}
\begin{equation}
\label{equ:comov-displacement}
	x^i(\vec\theta, \chi) = d_A(\chi)\theta^i - 2 \int_0^{\chi}\diff \chi' d_A(\chi-\chi')\Phi_{,i}(\vec x(\vec\theta, \chi'), \chi') \ ,
\end{equation}
where $\vec\theta$ represents the angle between the perturbed and fiducial ray at the observer. Vector quantities are denoted by lowercase Latin indices and partial derivatives with respect to comoving transverse coordinates, i.e., those perpendicular to the line-of-sight, are denoted by a comma. We make use of the sum convention where repeated indices are summed over. Unless otherwise noted, this sum only includes the two transverse directions. The Jacobi map is defined as the derivative of the deflection angle $\frac{\vec x(\vec\theta, \chi)}{d_A(\chi)}$ with respect to $\vec\theta$, i.e.,
\begin{splitequation}
	\label{equ:distortion-matrix}
	\cA_{ij}(\vec\theta, \chi) = \frac{\partial x^i(\vec\theta, \chi)}{d_A(\chi) \partial \theta^j} = \delta_{ij} - 2 \int_0^{\chi}\diff \chi' \frac{d_A(\chi-\chi')d_A(\chi')}{d_A(\chi)}\Phi_{,ik}(\vec x(\vec\theta, \chi'), \chi')\cA_{kj}(\vec\theta, \chi')  \ .
\end{splitequation}
It can be expressed in terms of the convergence $\kappa$, shear $\gamma_1$, $\gamma_2$, and rotation $\omega$ as
\begin{equation}
	\label{equ:psi-def}
	\cA_{ij} = 
	\begin{pmatrix}
		1 - \kappa - \gamma_1 	&-\gamma_2 - \omega \\
		-\gamma_2 + \omega		&1 - \kappa + \gamma_1
	\end{pmatrix} = \delta_{ij} - \psi_{ij}\ .
\end{equation}
Here we have introduced the distortion tensor $\psi_{ij}$ as a measure of the deviation from the identity map. The convergence is then given by the trace of the distortion tensor
\begin{splitequation}
\label{equ:kappa-def}
	\kappa = \frac{1}{2}\psi_{ii} \ .
\end{splitequation}
Using \eqref{equ:distortion-matrix} and \eqref{equ:kappa-def}, we find for the first order convergence
\begin{splitequation}
	\label{equ:kappa-singlesource}
	\kappa^{(1)}(\vec\theta, \chi_{_S}) = \int_0^{\chi_{_S}} \diff \chi' K(\chi_{_S},\chi')\Phi_{,ii}(d_A(\chi')\vec\theta,\chi') \ ,
\end{splitequation}
where we have defined the kernel
\begin{equation}
	K(\chi_{_S},\chi') = \frac{d_A(\chi_{_S}-\chi') d_A(\chi')}{d_A(\chi_{_S})}\Theta(\chi_{_S}-\chi') \ .
\end{equation}
Equation \eqref{equ:kappa-singlesource} describes the convergence due to a single source at a comoving distance $\chi_{_S}=\chi(z_{_S})$ from the observer. The convergence of a population of sources with redshift distribution $n(z)\diff z$ is found by averaging over the sources with $n(z)$ as the weighting factor. One then finds
\begin{splitequation}
	\kappa^{(1)}(\vec\theta) = \int_0^{\infty} \diff \chi(z) p(z)\frac{\diff z}{\diff \chi} \kappa^{(1)}(\vec\theta, \chi(z)) = \int_0^{\infty}\diff \chi W^\kappa(\chi)\Phi_{,ii}(d_A(\chi)\vec\theta, \chi) \ ,
\end{splitequation}
with the kernel given by
\begin{splitequation}
	W^\kappa(\chi) = \int_\chi^{\infty} \diff \chi' p(z)\frac{\diff z}{\diff \chi'}K(\chi', \chi) \ .
\end{splitequation}

The tSZ effect involves the inverse Compton scattering of CMB photons off relativistic electrons \cite{Sunyaev1972}. This introduces a frequency dependent temperature shift $\Delta T$ in the observed CMB temperature. The temperature shift at position $\vec\theta$ on the sky and frequency $\nu$ can be parameterized as
\begin{splitequation}
	\frac{\Delta T}{T_0}(\vec\theta, \nu) = y(\vec\theta) S_{SZ}(\nu) \ ,
\end{splitequation}
where the Compton $y(\vec\theta)$ parameter encodes the spatial and $S_{SZ}(\nu)$ the spectral dependence. The Compton $y$ parameter is defined as the line-of-sight integral over the electron pressure. In this work we adapt the constant bias model of \cite{VanWaerbeke2014} to simplify the analysis. It has been shown in \cite{Ma2014} that the constant bias model is consistent with a halo model approach, thus justifying the use of the simpler model. For a constant bias, the $y$ parameter can be written as an integral over the density contrast $\delta$, i.e.,
\begin{equation}
	y(\vec\theta) = \int_0^{\chi_H} \diff \chi \tilde W^y(\chi) \delta(\vec x(\vec \theta, \chi),\chi) \ ,
\end{equation}
where $\chi_H$ is the comoving distance to the surface of last scattering.
We express the density contrast in terms of the Newtonian potential through the Poisson equation as
\begin{equation}
	\label{equ:y-def}
	y(\vec\theta) = \int_0^{\chi_H} \diff \chi \tilde W^y(\chi) \frac{2 a \Delta \Phi(\vec x(\vec \theta, \chi),\chi)}{3 H_0^2 \Omega_m} = \int_0^{\chi_H} \diff \chi W^y(\chi) \Delta \Phi(\vec x(\vec \theta, \chi),\chi) \ ,
\end{equation}
where we have absorbed the factors from the Poisson equation in the new kernel $W^y(\chi)$.

Ultimately, we are interested in the angular cross power spectrum $C^{y\kappa}_\ell$. In this work we assume that the convergence is derived from shear measurements of galaxy surveys. The sky coverage of these surveys is still relatively small, although this will not be the case of future surveys such as Euclid and LSST, allowing the analysis to proceed in the flat-sky approximation. 
Using the definition of the 2d Fourier transform \eqref{equ:def-angular-fourier-transfrom}, we can write the angular cross power spectrum of $\hat y(\vec\ell)$ and $\hat \kappa(\vec\ell)$ as
\begin{splitequation}
	\label{equ:y-kappa-1st-verbose}
		\braket{\hat y^{(1)}(\vec\ell_1)\hat\kappa^{(1)}(\vec\ell_2)} &= (2\pi)^2 \ddelta^2(\vec\ell_1 + \vec\ell_2) C^{(2)}_{\ell_1} \\
		&= \int_0^{\chi_H}\diff \chi \diff \chi' W^y(\chi) W^\kappa(\chi')\frac{\abs{\vec\ell_1}^2}{d_A(\chi)^2}\frac{\abs{\vec\ell_2}^2}{d_A(\chi')^2} \braket{\hat\phi(\vec\ell_1,\chi)\hat\phi(\vec\ell_2,\chi')} \ ,
\end{splitequation}
where we have dropped the contributions of the derivatives along the line-of-sight in \eqref{equ:y-def}. Under the Limber approximation \cite{Limber1953, Kaiser1992} one assumes that the transverse modes are much large than the longitudinal modes, i.e., $\frac{\ell}{d_A(\chi)} \gg k_3$. This ceases to be true on large scales, where extensions to the Limber approximation such as \cite{LoVerde2008} or an exact full-sky treatment have to be employed. For the scales of interest in this work the Limber approximation is sufficient though. In fact, it has been shown in \cite{Bernardeau2012} that the lowest order Limber approximation is an excellent fit down to multipoles of $\ell \sim 20$. Expressing the two-point function in \eqref{equ:y-kappa-1st-verbose} in terms of the power spectrum \eqref{equ:powerspectrum-def} we find for the $y$-$\kappa$ cross power spectrum
\begin{splitequation}
	\label{equ:y-kappa-1st}
		\braket{\hat y^{(1)}(\vec\ell_1)\hat\kappa^{(1)}(\vec\ell_2)}  &= (2\pi)^2\ddelta^2\left(\vec\ell_1 + \vec\ell_2\right) \int_0^{\chi_H}\diff \chi W^y(\chi)W^\kappa(\chi) \frac{ \abs{\vec\ell_1}^4}{d_A(\chi)^4}\Pp\left(\frac{\abs{\vec\ell_1}}{d_A(\chi)}, \chi \right) \ .
\end{splitequation}
Note that upon replacing the kernel for the $y$ parameter $W^y$ with $W^\kappa$, one recovers the well known expression for the angular power spectrum of the convergence.

\section{Corrections}
\label{sec:correction}
To consistently treat fourth order corrections to the cross spectrum we need to include terms up to third order in $\Phi$ of the Compton $y$ parameter and convergence $\kappa$ \cite{Cooray2002}. Indeed, it has been shown in \cite{Krause2010} that divergences in second-second order cross terms cancel with corresponding divergences in first-third order cross terms. It is thus important to find expressions for the two fields $y$ and $\kappa$ up to third order. A full sky treatment of lensing observables to even second order is already a formidable task \cite{Bernardeau2010, Bernardeau2012, Vanderveld2011}; a full sky derivation to third order would be beyond the scope of this work. Fortunately, the calculations can be simplified greatly by restricting ourselves to small scales. We follow \cite{Dodelson2005} to identify the terms that contribute dominantly at small scales and those that can be neglected.

Broadly speaking, on small scales terms with the most angular derivatives are expected to dominate. At second order this are the well known Born approximation, lens-lens coupling, and reduced shear contributions \cite{Bernardeau1997,Schneider1998, Cooray2002, Dodelson2005, Dodelson2006, Shapiro2009, Krause2010}. Third order terms derived from the aforementioned have at least the same number of angular derivatives and are therefore expected to be the dominant third order contributions. We discuss these contributions in section~\ref{sec:born-lens-lens} and section~\ref{sec:reduced-shear}.

Recent work by \cite{Andrianomena2014} found that contributions from peculiar velocities to the convergence can be as large as the primary contribution from scalar modes in certain redshift ranges. Even though peculiar velocities formally affect the convergence at first order, they affect the shear only at second order \cite{Bonvin2008}. In the case where the convergence is derived from shear measurements, as we assume in this work, the effect of peculiar velocities enters only at second order. We investigate the effect of peculiar velocities in section~\ref{sec:redshift-distortions}. Vector modes induced by second order perturbations have been shown to yield corrections of similar magnitude as traditional Born and lens-lens terms \cite{Andrianomena2014}. We show that vector modes do not contribute to the $y$-$\kappa$ cross spectrum at fourth order in section~\ref{sec:vector-modes}.

To distinguish the different corrections to the convergence we denote them by subscripts; $\kappa_{std}$ refers to corrections due to Born approximation and lens-lens coupling, $\kappa_{rs}$ to corrections due to the reduced shear, $\kappa_z$, and $\kappa_v$ to corrections due to redshift distortions and vector modes, respectively.

\subsection{Born approximation and lens-lens coupling}
\label{sec:born-lens-lens}
For the derivation of the Born and lens-lens coupling terms we roughly follow \cite{Krause2010}, in that we expand the solution to the geodesic deviation equation \eqref{equ:comov-displacement} systematically in powers of $\Phi$. Alternatively, one could expand the terms in the distortion matrix \eqref{equ:distortion-matrix}, which makes the physical meaning of the terms more apparent.

We expand the comoving transverse displacement in powers of the potential $\Phi$ as
\begin{equation}
	\vec x = \vec x^{(0)} + \vec x^{(1)} + \vec x^{(2)} + \vec x^{(3)} + \cO(\Phi^4) \ ,
\end{equation}
where the superscript in parentheses denotes the order of the expansion. The zeroth and first order contributions are given by
\begin{splitequation}
	\vec x^{(0)} = d_A(\chi)\vec\theta ,\quad x^{(1)i}(\vec\theta, \chi) = - 2 \int_0^\chi\diff \chi' d_A(\chi-\chi')\Phi_{,i}(d_A(\chi')\vec\theta, \chi') \ .
\end{splitequation}
The higher order contributions can be found by Taylor expanding $\Phi(\vec x)$ in \eqref{equ:comov-displacement} around the zeroth order solution $\vec x^{(0)}(\vec\theta,\chi) = d_A(\chi)\vec\theta$. The potential can then be expanded as $\Phi = \Phi^{(1)} + \Phi^{(2)} + \Phi^{(3)} + \cO(\Phi^4)$, with 
\begin{splitequation}
	\label{equ:phi-expansion}
	\Phi^{(1)}(\vec x) &= \Phi(\vec x^{(0)}) \\
	\Phi^{(2)}(\vec x) &= \Phi_{,i}(\vec x^{(0)}) x^{(1) i}\\
	\Phi^{(3)}(\vec x) &= \frac{1}{2}\Phi_{,ij}(\vec x^{(0)}) x^{(1) i}x^{(1) j} +  \Phi_{,i}(\vec x^{(0)}) x^{(2) i} \ .
\end{splitequation}
By replacing the $\Phi$ with $\Phi^{(2)}$ in \eqref{equ:comov-displacement}, we can write the second order deflection angle as
\begin{splitequation}
	\frac{x^{(2)i}(\vec\theta, \chi)}{d_A(\chi)} &= - 2 \int_0^\chi\diff \chi' \frac{d_A(\chi-\chi')}{d_A(\chi)}\Phi^{(2)}_{,i}(\vec x(\vec\theta, \chi'),\chi') \\
					 &= 4  \int_0^\chi\diff \chi' \int_0^{\chi'}\diff \chi'' K(\chi,\chi')\frac{K(\chi',\chi'')}{d_A(\chi'')}\Phi_{,ij}(\chi') \Phi_{,j}(\chi'') \ ,
\end{splitequation}
where we have dropped the angular dependence of the potentials for brevity. We adapt this shorthand for the rest of this work, i.e., unless otherwise noted $\Phi(\vec x^{(0)}(\vec\theta,\chi), \chi)$ is written as $\Phi(\chi)$. Analogously, the third order deflection angle can be written as
\begin{splitequation}
	\frac{x^{(3)i}(\vec\theta, \chi)}{d_A(\chi)} 
					 &= -4  \int_0^\chi\diff \chi' \int_0^{\chi'}\diff \chi'' \int_0^{\chi'}\diff \chi''' K(\chi,\chi')\frac{K(\chi',\chi'')K(\chi',\chi''')d_A(\chi')}{d_A(\chi'')d_A(\chi''')} \\
					 &\qquad\qquad \times \Phi_{,ijk}(\chi') \Phi_{,j}(\chi'') \Phi_{,k}(\chi''') \\
					 &\quad -8  \int_0^\chi\diff \chi' \int_0^{\chi'}\diff \chi'' \int_0^{\chi''}\diff \chi''' K(\chi,\chi')K(\chi',\chi'')\frac{K(\chi'',\chi''')}{d_A(\chi''')} \\
					 &\qquad\qquad \times \Phi_{,ij}(\chi') \Phi_{,jk}(\chi'') \Phi_{,k}(\chi''') \ .
\end{splitequation}

\subsubsection{Convergence}
Equipped with second and third order expressions for the deflection angle it is straightforward to derive expressions for the convergence. Using the relation of the convergence to the trace of the distortion tensor \eqref{equ:kappa-def}, we can readily write down the second and third order expressions for the convergence. At second order this is
\begin{splitequation}
	\kappa^{(2)}_{std}(\vec\theta, \chi{_S}) 
				&= -2\int_0^{\chi_S}\diff \chi' \int_0^{\chi'}\diff \chi'' K(\chi,\chi')K(\chi',\chi'') \\
				&\qquad\qquad \times \left(\frac{d_A(\chi')}{d_A(\chi'')}\Phi_{,iij}(\chi') \Phi_{,j}(\chi'') + \Phi_{,ij}(\chi') \Phi_{,ji}(\chi'')\right) \ .
\end{splitequation}
The first term in the bracket is the well known Born term, while the second is the lens-lens coupling term. The extra factors of the comoving angular distance $d_A(\chi)$ arise because the derivative in \eqref{equ:distortion-matrix} is with respect to the angular deviation $\vec\theta$, whereas the potential is a function of the comoving transverse distance $\vec x^{(0)}=d_A(\chi)\vec\theta$.

The third order expression for the convergence is analogously found to be
\begin{splitequation}
	\label{equ:kappa-3rd-order}
	\kappa^{(3)}_{std}(\vec\theta, \chi_S) 
				&= 2\int_0^{\chi_S}\diff \chi' \int_0^{\chi'}\diff \chi'' \int_0^{\chi'}\diff \chi''' K(\chi,\chi')K(\chi',\chi'')K(\chi',\chi''')  \\
					 &\qquad\qquad \times\left(\frac{d_A(\chi')^2}{d_A(\chi'')d_A(\chi''')}\Phi_{,iijk}(\chi')\Phi_{,j}(\chi'')\Phi_{,k}(\chi''')\right) \\
				&\ +4\int_0^{\chi_S}\diff \chi' \int_0^{\chi'}\diff \chi'' \int_0^{\chi'}\diff \chi''' K(\chi,\chi')K(\chi',\chi'')K(\chi',\chi''')  \\
					 &\qquad\qquad \times\left(\frac{d_A(\chi')}{d_A(\chi''')}\Phi_{,ijk}(\chi')\Phi_{,ji}(\chi'')\Phi_{,k}(\chi''')\right) \\
				&\ +4\int_0^{\chi_S}\diff \chi' \int_0^{\chi'}\diff \chi'' \int_0^{\chi''}\diff \chi''' K(\chi,\chi')K(\chi',\chi'')\frac{K(\chi'',\chi''')}{d_A(\chi''')} \\
					 &\qquad\qquad \times\frac{\partial}{\partial \theta^i}\left(\Phi_{,ij}(\chi')\Phi_{,jk}(\chi'')\Phi_{,k}(\chi''')\right) \ .
\end{splitequation}
The term on line 2 corresponds a second order Born correction, the term on line 4 to a mixed Born-lens-coupling, and the three terms on line 6 to a second order Born correction, Born-lens-coupling, second order lens-lens coupling, respectively.

\subsubsection{Compton $y$ parameter}
The second and third order contributions to the Compton $y$ parameter are somewhat easier to derive, as there are no lens-lens coupling terms. As in the case of the convergence, we replace $\Phi$ in  \eqref{equ:y-def} by its expansion \eqref{equ:phi-expansion}. The second order contribution to the $y$ parameter is then
\begin{equation}
	\begin{split}
		y^{(2)}(\vec\theta) &= -2\int_0^{\chi_H}\diff \chi' \int_0^{\chi'}\diff \chi'' W^y(\chi')K(\chi',\chi'')\frac{d_A(\chi')}{d_A(\chi'')}\Phi_{,iij}(\chi')\Phi_{,j}(\chi'') \ .
	\end{split}
\end{equation}
The third order term follows analogously and is given by
\begin{equation}
	\label{equ:y-3rd-order}
	\begin{split}
		y^{(3)}(\vec\theta) &= 2\int_0^{\chi_H}\diff \chi' \int_0^{\chi'}\diff \chi'' \int_0^{\chi'}\diff \chi''' W^y(\chi')K(\chi',\chi'')K(\chi',\chi''')\frac{d_A(\chi')^2}{d_A(\chi'')d_A(\chi''')}  \\
					 &\qquad\qquad \times\Phi_{,iijk}(\chi')\Phi_{,j}(\chi'')\Phi_{,k}(\chi''') \\
		&\quad + 4\int_0^{\chi_H}\diff \chi' \int_0^{\chi'}\diff \chi'' \int_0^{\chi''}\diff \chi''' W^y(\chi')K(\chi',\chi'')K(\chi'',\chi''')\frac{d_A(\chi')}{d_A(\chi''')} \\
					 &\qquad\qquad \times\Phi_{,iij}(\chi')\Phi_{,jk}(\chi'')\Phi_{,k}(\chi''') \ .
	\end{split}
\end{equation}
Both terms are due to the Born approximation. The term on the second line stems from the $\frac{1}{2}\Phi_{,ij}(\vec x^{(0)}) x^{(1) i}x^{(1) j}$ term in the third order contribution to $\Phi$ in \eqref{equ:phi-expansion}, whereas the term on line 4 in \eqref{equ:y-3rd-order} is due to the $\Phi_{,i}(\vec x^{(0)}) x^{(2) i}$  term in \eqref{equ:phi-expansion}.

\subsubsection{Cross correlations}
The second-second order contribution to the angular $y$-$\kappa$ cross power spectrum due to Born and lens-lens terms can be derived by taking the ensemble average of the product of the Fourier space expressions $\hat y^{(2)}(\vec\ell_1)$ and $\hat \kappa^{(2)}(\vec\ell_2)$. Using the results from appendix~\ref{sec:fourier-identities}, we find
\begin{splitequation}
	\braket{\hat y^{(2)}(\vec\ell_1)\hat\kappa^{(2)}_{std}(\vec\ell_2)} &= 4\int_0^{\chi_H} \diff \chi_y \diff \chi_\kappa \int_0^{\chi_y}\diff \chi_y' \int_0^{\chi_\kappa}\diff \chi_\kappa' \frac{W^y(\chi_y)W^\kappa(\chi_\kappa)K(\chi_y,\chi_y')K(\chi_\kappa,\chi_\kappa')}{d_A(\chi_y)^2 d_A(\chi_y')^2 d_A(\chi_\kappa)^2 d_A(\chi_\kappa')^2}  \\
		&\quad\times \int \frac{\diff^2 \vec {\ell'} \diff^2 \vec {\ell''}}{(2\pi)^4} \abs{\vec {\ell'}}^2 \vec {\ell'}(\vec\ell_1 - \vec {\ell'})\left[ \abs{\vec {\ell''}}^2 \vec {\ell''}(\vec\ell_2 - \vec {\ell''})+ \left(\vec {\ell''} (\vec\ell_2 - \vec {\ell''})\right)^2\right] \\
		&\quad\times \braket{\hat\phi(\vec {\ell'}, \chi_y)\hat\phi(\vec\ell_1 - \vec {\ell'}, \chi_y')\hat\phi(\vec {\ell''}, \chi_\kappa)\hat\phi(\vec\ell_2 - \vec {\ell''}, \chi_\kappa')} \ ,
\end{splitequation}
where we used the kernel $W^\kappa$ for a source distribution $n(z)$ instead of a single source at redshift $z_S$.
The four-point function on the last line is made up of one connected and three unconnected terms. The connected term is proportional to the trispectrum \eqref{equ:trispectrum-def}. Under the Limber approximation this introduces a product of delta functions $\ddelta(\chi_y - \chi_y')\ddelta(\chi_y - \chi_\kappa)\ddelta(\chi_y - \chi_\kappa')$, setting all comoving distances along the line-of-sight equal. The kernel $K(\chi,\chi')$ is zero for $\chi\leq \chi'$, thus eliminating the contribution from the connected part of the correlation function. The unconnected part can be decomposed into three products of two-point functions by Wick's theorem. Each of the two-point functions yields a delta function times a power spectrum. The term proportional to $\ddelta(\chi_y - \chi_y')\ddelta(\chi_\kappa - \chi_\kappa')$ is zero because $K(\chi,\chi)=0$. The term proportional to $\ddelta(\chi_y - \chi_\kappa')\ddelta(\chi_y' - \chi_\kappa)$ is zero because $K(\chi, \chi')K(\chi',\chi)\equiv 0$.
The only surviving term is proportional to $\ddelta(\chi_y - \chi_\kappa)\ddelta(\chi_y' - \chi_\kappa')$, and upon evaluating the integrals gives for the angular cross power spectrum
\begin{splitequation}
	\label{equ:y2-kappa2}
	\braket{\hat y^{(2)}(\vec\ell_1)\hat\kappa^{(2)}_{std}(\vec\ell_2)} &= 4 (2\pi)^2\ddelta^2(\vec\ell_1+\vec\ell_2) \int_0^{\chi_H} \diff \chi \int_0^{\chi}\diff \chi' \frac{W^y(\chi)W^\kappa(\chi)K(\chi,\chi')^2}{d_A(\chi)^6 d_A(\chi')^6} \int \frac{\diff^2 \vec {\ell'}}{(2\pi)^2} \\
		&\quad\times \abs{\vec {\ell'}}^2 \vec\ell_1\vec{\ell'}\left(\vec {\ell'}(\vec\ell_1 - \vec {\ell'})\right)^2  \Pp\left(\frac{\abs{\vec {\ell'}}}{d_A(\chi)},\chi\right)\Pp\left(\frac{\abs{\vec\ell_1 - \vec {\ell'}}}{d_A(\chi')},\chi'\right) \ .
\end{splitequation}
The derivation for the first-third order contributions proceeds similarly. The connected correlation function drops out for the same reason as in the second-second order case. Furthermore, terms in the third order expressions for $y$ and $\kappa$ that include a line-of-sight kernel proportional to $K(\chi',\chi'')K(\chi'',\chi''')$, i.e., line 5 in \eqref{equ:kappa-3rd-order} and line 3 in \eqref{equ:y-3rd-order}, do not contribute to the power spectra because the kernel is zero for all possible contractions of the correlation function. 

The contribution from the second term in \eqref{equ:kappa-3rd-order}, i.e., line 4, to the cross power spectrum is proportional to 
\begin{equation*}
	\int \frac{\diff^2 \vec {\ell'}}{(2\pi)^2} \left(\vec {\ell_1}\vec {\ell'}\right)^3 \Pp\left(\frac{\abs{\vec\ell_1}}{d_A(\chi)},\chi\right)\Pp\left(\frac{\abs{\vec {\ell'}}}{d_A(\chi')},\chi'\right) \ ,
\end{equation*}
which is zero due to the antisymmetry of the integral under the transformation $\vec{\ell'} \rightarrow -\vec{\ell'}$ \cite{Krause2010}. Hence, only the first Born term in \eqref{equ:kappa-3rd-order} contributes to $\braket{\hat y^{(1)}\hat\kappa^{(3)}}$. We find
\begin{splitequation}
	\label{equ:y1-kappa3}
	\braket{\hat y^{(1)}(\vec\ell_1)\hat\kappa^{(3)}_{std}(\vec\ell_2)} &= -2 (2\pi)^2\ddelta^2(\vec\ell_1+\vec\ell_2) \int_0^{\chi_H} \diff \chi \int_0^{\chi}\diff \chi' \frac{W^y(\chi)W^\kappa(\chi)K(\chi,\chi')^2}{d_A(\chi)^6 d_A(\chi')^6}\\
		&\quad\times  \int \frac{\diff^2 \vec {\ell'}}{(2\pi)^2}\ \abs{\vec\ell_1}^4 \left(\vec\ell_1 \vec {\ell'}\right)^2 \Pp\left(\frac{\abs{\vec\ell_1}}{d_A(\chi)},\chi\right)\Pp\left(\frac{\abs{\vec {\ell'}}}{d_A(\chi')},\chi'\right) \ .
\end{splitequation}
Since the only contribution to $\braket{\hat y^{(3)}\hat\kappa^{(1)}}$ comes from the first term in \eqref{equ:y-3rd-order}, which is identical to the first term in \eqref{equ:kappa-3rd-order} up to an interchange of the kernels $W^y(\chi)$ and $W^\kappa(\chi)$, the cross power spectra $\braket{\hat y^{(3)}\hat\kappa^{(1)}}$  and $\braket{\hat y^{(1)}\hat\kappa^{(3)}}$ are identical.

\subsection{Reduced shear}
\label{sec:reduced-shear}
At first order the shear and convergence are related by
\begin{equation}
	\label{equ:kappa-gamma-rel}
	\hat\kappa(\vec\ell) = T^I(\vec\ell) \hat\gamma_I(\vec\ell) \ ,\quad T^1(\vec\ell) = \cos{2\phi_\ell}\ ,\quad T^2(\vec\ell) = \sin{2\phi_\ell} \ ,
\end{equation}
where $\phi_\ell$ is the angle between the two-dimensional wave-vector $\vec\ell$ and some fixed reference axis. The components of the shear and other polar quantities are labeled by capital Latin indices. It can be shown that this relation holds exactly up to second order and under the Limber approximation up to third order; see appendix \ref{sec:gamma-kappa-rel} for details.
In the weak lensing regime, the measured quantity is not the shear itself but the reduced shear, conventionally defined as
\begin{equation}
	\label{equ:g-sym-def}
	g_I = \frac{\gamma_I}{1-\kappa} \ , \quad I=1,2 \ .
\end{equation}
The definition \eqref{equ:g-sym-def} of the reduced shear is based on the assumption that the Jacobi map \eqref{equ:distortion-matrix} is symmetric. In general the Jacobi map is not symmetric however, because lens-lens couplings generate the anti-symmetric contribution $\omega$. Including the anti-symmetric terms in the Jacobi map, the generalized reduced shear in complex notation is given by (see appendix \ref{sec:rotation})
\begin{equation}
	g=\frac{\gamma_1+\imath\gamma_2}{1-\kappa+\imath\omega} \ .
\end{equation}
Accounting for the reduced shear in the relation \eqref{equ:kappa-gamma-rel} amounts to replacing the shear $\gamma_I$ with the reduced shear $g_I$. Using \eqref{equ:general-g-approx} and expanding systematically in $\Phi$ to third order we can express the observed convergence as
\begin{splitequation}
	\label{equ:kappa-gamma-full}
	\hat\kappa_{obs} = T^I\hat g_I = T^I &\left(\hat\gamma_I^{(1)} + (\hat\gamma^{(2)}_{std})_I + (\hat\gamma^{(3)}_{std})_I \right.\\
					&+ \hat\gamma_I^{(1)}*\hat\kappa^{(1)} + \hat\gamma_I^{(1)}*\hat\kappa^{(1)}*\hat\kappa^{(1)} + \hat\gamma_I^{(1)}*\hat\kappa^{(2)}_{std} + (\hat\gamma^{(2)}_{std})_I*\hat\kappa^{(1)} \\
					&\left. + R(\hat\omega^{(2)}_{std})_{IJ}*\hat\gamma_J^{(1)} \right) + \cO(\Phi^4) \ ,
\end{splitequation}
where $*$ stands for a convolution in Fourier space.
As shown in appendix \ref{sec:gamma-kappa-rel}, the first line is equivalent to $\hat\kappa^{(1)} + \hat\kappa^{(2)}_{std} + \hat\kappa^{(3)}_{std}$, where $\hat\kappa^{(2)}_{std}$ and $\hat\kappa^{(3)}_{std}$ denote the corrections due to Born approximation and lens-lens coupling. The second line includes the well known contributions from the reduced shear \cite{Schneider1998, Dodelson2006, Shapiro2009, Krause2010}, while the third line is a novel contribution due to second order induced rotations. The sole second order correction due to reduced shear to the convergence is 
\begin{splitequation}
\label{equ:kappa-2nd-rs}
	\hat\kappa^{(2)}_{rs}(\vec\ell) &= [T^I(\vec\ell)\hat\gamma^{(1)}_I * \hat\kappa^{(1)}](\vec\ell)= \int \frac{\diff^2 \vec {\ell'}}{(2\pi)^2} T^I(\vec\ell)\hat\gamma^{(1)}_I(\vec  {\ell'})\hat\kappa^{(1)}(\vec \ell - \vec  {\ell'}) \\
		&= \int \frac{\diff^2 \vec  {\ell'}}{(2\pi)^2} \cos(2\phi_ {\ell'} - 2\phi_\ell) \hat\kappa^{(1)}(\vec  {\ell'})\hat\kappa^{(1)}(\vec \ell - \vec  {\ell'}) \ ,
\end{splitequation}
where we used the identity $T^I(\vec\ell)T_I(\vec {\ell'}) = \cos(2\phi_{\ell'} - 2\phi_\ell)$.
Since the reduced shear is an intrinsic lensing effect, it does not affect the Compton $y$ parameter. The lowest order contribution to the cross power spectrum is therefore formed by the first order $y$ parameter \eqref{equ:y-def} and second order reduced shear correction \eqref{equ:kappa-2nd-rs}, i.e.,
\begin{splitequation}
	\label{equ:3rd-order-rs}
	\braket{\hat y^{(1)}(\vec\ell_1)\hat\kappa^{(2)}_{rs}(\vec\ell_2)} &= - (2\pi)^2 \ddelta^2(\vec\ell_1 + \vec\ell_2)\int \diff \chi \frac{W^y(\chi)\left(W^\kappa(\chi)\right)^2}{d_A(\chi)^{10}} \int \frac{\diff^2 \vec {\ell'}}{(2\pi)^2} \\
	&\quad \times  \abs{\vec\ell_1}^2 \abs{\vec {\ell'}}^2\abs{\vec\ell_2 - \vec {\ell'}}^2 \cos(2\phi_{\ell'} - 2\phi_{\ell_2}) \cB_\Phi\left(\frac{\abs{\vec \ell_1}}{d_A(\chi)},\frac{\abs{\vec {\ell'}}}{d_A(\chi)},\frac{\abs{\vec\ell_2 - \vec {\ell'}}}{d_A(\chi)}\right) \ ,
\end{splitequation}
where we used the definition \eqref{equ:bispectrum-def} of the bispectrum. Unlike in the case of the Born and lens-lens terms, there is a third order contribution to the cross power spectrum. 
As a consistency check, one can show that upon replacing $W^y$ by $W^\kappa$, and using the fact that to first order the convergence is the same as the E-mode of the shear, one recovers the expression for the correction to the E-mode shear due to reduced shear in \cite{Dodelson2006}.

To analyze the first-third order contributions, we split the third order contribution to the convergence due to the reduced shear in \eqref{equ:kappa-gamma-full} into three components
\begin{subequations}
	\label{equ:kappa-3rd-RS-split}
	\begin{align}
		\hat\kappa^{(3,A)}_{rs}(\vec\ell) &= T^I(\vec\ell)[\hat\gamma^{(1)}_I * \hat\kappa^{(1)}* \hat\kappa^{(1)} ](\vec\ell) \label{euq:kappa-3rd-RS-A} \ ,\\
		\hat\kappa^{(3,B)}_{rs}(\vec\ell) &= T^I(\vec\ell)[\hat\gamma^{(1)}_I * \hat\kappa^{(2)}_{std} +  (\hat\gamma^{(2)}_{std})_I * \hat\kappa^{(1)} ](\vec\ell) \label{euq:kappa-3rd-RS-B} \ ,\\
		\hat\kappa^{(3,C)}_{rs}(\vec\ell) &=T^I(\vec\ell) [R(\hat\omega^{(2)}_{std})_{IJ}*\hat\gamma_J^{(1)} ](\vec\ell) \label{euq:kappa-3rd-RS-C}\ .
	\end{align}
\end{subequations}
The cross power spectrum of $\hat\kappa^{(3,A)}_{rs}$ with $y$ is then
\begin{splitequation}
	\label{equ:y-kappa-3-RS-A-verbose}
	\braket{\hat y^{(1)}(\vec\ell_1)\hat\kappa^{(3,A)}_{rs}(\vec\ell_2)} &= \int \diff \chi_y \frac{W^y(\chi_y)}{d_A(\chi_y)^2} \prod_{i=1}^3\diff \chi_i \frac{W^\kappa(\chi_i)}{d_A(\chi_i)^2} \int \frac{\diff^2 \vec {\ell'}\diff^2\vec{\ell''}}{(2\pi)^4} \cos(2\phi_{\ell_2} - 2\phi_{\ell'}) \\
	&\quad\times \abs{\vec\ell_1}^2 \abs{\vec{\ell'}}^2 \abs{\vec{\ell''}}^2 \abs{\vec \ell_2 - \vec {\ell'} - \vec{\ell''}}^2 \\
	&\quad\times \braket{\hat\phi(\vec\ell_1, \chi_y)\hat\phi(\vec{\ell'}, \chi_1)\hat\phi(\vec{\ell''}, \chi_2)\hat\phi(\vec \ell_2 - \vec {\ell'} - \vec{\ell''}, \chi_3)} \ .
\end{splitequation}
Because the line-of-sight integral does not include the kernel $K(\chi,\chi')$, like in the case of the third order cross power spectrum \eqref{equ:3rd-order-rs}, the connected part of the four-point function does not vanish. The connected  and unconnected contributions to the cross power spectrum are found to be
\begin{subequations}
	\label{equ:y-kappa-3-RS-A}
	\begin{align}
		\begin{split}
			\braket{\hat y^{(1)}(\vec\ell_1)\hat\kappa^{(3,A)}_{rs}(\vec\ell_2)}_c &= (2\pi)^2\ddelta^2\left(\vec\ell_1 + \vec\ell_2\right)\int \diff \chi \frac{W^y(\chi) W^\kappa(\chi)^3}{d_A(\chi)^{14}} \int \frac{\diff^2 \vec {\ell'}\diff^2\vec{\ell''}}{(2\pi)^4} \cos(2\phi_{\ell_2} - 2\phi_{\ell'}) \\
					&\quad\times \abs{\vec\ell_1}^2\abs{\vec{\ell'}}^2\abs{\vec{\ell''}}^2\abs{\vec \ell_1 + \vec {\ell'} + \vec{\ell''}}^2 \cT_\Phi\left(\frac{\vec\ell_1}{d_A(\chi)},\frac{\vec{\ell'}}{d_A(\chi)},\frac{\vec{\ell''}}{d_A(\chi)},-\frac{\vec \ell_1 + \vec {\ell'} + \vec{\ell''}}{d_A(\chi)},\chi\right)
		\end{split} \\
		\braket{\hat y^{(1)}(\vec\ell_1)\hat\kappa^{(3,A)}_{rs}(\vec\ell_2)}_g &= \braket{\hat y^{(1)}(\vec\ell_1)\hat\kappa^{(1)}(\vec\ell_2)}\int \frac{\diff^2\vec{\ell'}}{(2\pi)^2}C^{\kappa\kappa, 2}_{\ell'} \ ,
	\end{align}
\end{subequations}
where the connected and unconnected parts are denoted by the subscript c and g, respectively. For the derivation of the connected part we have used the definition of the trispectrum \ref{equ:trispectrum-def}.
Replacing the kernel $W^y$ by $W^\kappa$ we recover again the same expression as found in \cite{Krause2010}.

The two other third order contributions to the convergence due the reduced shear \eqref{euq:kappa-3rd-RS-B} and \eqref{euq:kappa-3rd-RS-C} involve second order Born and lens-lens corrections, i.e., include the coupling kernel $K(\chi,\chi')$ in their line-of-sight integrals. Hence, only their unconnected parts contribute to the cross power spectrum. The derivation proceeds as for the other terms discussed so far, albeit with somewhat more complicated expressions, as there are now two contractions of the four-point function that survive. The contribution involving $\hat\kappa^{(3,B)}_{rs}$ is
\begin{splitequation}
	\label{equ:y1-kappa3-RS-B}
	\braket{\hat y^{(1)}(\vec\ell_1)\hat\kappa^{(3,B)}_{rs}(\vec\ell_2)} &= -2(2\pi)^2\ddelta^2\left(\vec\ell_1 + \vec\ell_2\right)\int \diff \chi \diff \chi' \frac{W^y(\chi) W^\kappa(\chi')}{d_A(\chi)^6 d_A({\chi'})^6} \int \frac{\diff^2 \vec {\ell'}}{(2\pi)^2} \\
	&\quad\times \abs{\vec\ell_1}^2\abs{\vec{\ell'}}^2 \vec\ell_1\vec{\ell'} \Pp\left(\frac{\abs{\vec\ell_1}}{d_A(\chi)},\chi\right)\Pp\left(\frac{\abs{\vec{\ell'}}}{d_A(\chi')},\chi'\right)\\
	&\quad\times \left\{ \cos(2\phi_{\ell_2} - 2\phi_{\ell'})  \vec\ell_1\vec{\ell'} \left[W^\kappa(\chi)K(\chi,\chi')+W^\kappa(\chi')K(\chi',\chi)\right] \right. \\
	&\qquad \left. \cos(2\phi_{\ell_2} - 2\phi_{\vec{\ell'}+\vec\ell_1}) \left[\vec\ell_1(\vec{\ell'}+\vec\ell_1)W^\kappa(\chi)K(\chi,\chi')+\vec{\ell'}(\vec{\ell'}+\vec\ell_1)W^\kappa(\chi')K(\chi',\chi)\right] \right\} \ .
\end{splitequation}
The azimuthal integral of the third line can be done analytically and is equal to $\frac{\pi}{2}$. Note that our result differs from that obtained in \cite{Krause2010} by an extra factor of $\cos(2\phi_{\ell_2} - 2\phi_{\vec{\ell'}+\vec\ell_1})$ on fourth line.

Using the definition of the matrix $R(\omega)$ in \eqref{equ:R-omega}, the contribution $\hat\kappa^{(3,C)}_{rs}$ can be written as
\begin{splitequation}
	\hat\kappa^{(3,C)}_{rs}(\vec\ell) &=
	T^1(\vec\ell)[\gamma_2^{(1)}*\hat\omega^{(2)}_{std} ](\vec\ell) - T^2(\vec\ell)[\gamma_1^{(1)} * \hat\omega^{(2)}_{std} ](\vec\ell) \\
	&= \int \frac{\diff^2\vec{\ell'}}{(2\pi)^2} \sin(2\phi_{\ell'} - 2\phi_\ell)\hat\kappa^{(1)} (\vec{\ell'})\hat\omega^{(2)}_{std}(\vec\ell-\vec{\ell'}) \ .
\end{splitequation}
The cross power spectrum is therefore
\begin{splitequation}
	\label{equ:y1-kappa3-RS-C}
	\braket{\hat y^{(1)}(\vec\ell_1)\hat\kappa^{(3,C)}_{rs}(\vec\ell_2)} &= -2(2\pi)^2\ddelta^2\left(\vec\ell_1 + \vec\ell_2\right)\int \diff \chi \diff \chi' \frac{W^y(\chi) W^\kappa(\chi')}{d_A(\chi)^6 d_A({\chi'})^6} \int \frac{\diff^2 \vec {\ell'}}{(2\pi)^2} \\
	&\quad\times \abs{\vec\ell_1}^3\abs{\vec{\ell'}}^3 \vec\ell_1\vec{\ell'}\sin(\phi_{\ell_2}-\phi_{\ell'})\sin(2\phi_{\ell'} - 2\phi_{\ell_2}) \Pp\left(\frac{\abs{\vec\ell_1}}{d_A(\chi)},\chi\right)\Pp\left(\frac{\abs{\vec{\ell'}}}{d_A(\chi')},\chi'\right)\\
	&\quad\times  \left[W^\kappa(\chi)K(\chi,\chi')+W^\kappa(\chi')K(\chi',\chi)\right] \ .
\end{splitequation}
The mode coupling term can be reduced to $\abs{\vec\ell_1}^4\abs{\vec{\ell'}}^4\frac{\sin(2\phi_{\ell'} - 2\phi_{\ell_2})^2}{2}$. The azimuthal integral then evaluates to $\frac{\pi}{2}$. The contributions from $\hat\kappa^{(3,C)}_{rs}$ and from the first term in $\hat\kappa^{(3,B)}_{rs}$ are therefore identical.

Finally, we find the only second-second order contribution due to the reduced shear to the cross power spectrum to be
\begin{splitequation}
	\label{equ:y2-kappa2-RS}
	\braket{\hat y^{(2)}(\vec\ell_1)\hat\kappa^{(2)}_{rs}(\vec\ell_2)} &= -2(2\pi)^2\ddelta^2\left(\vec\ell_1 + \vec\ell_2\right)\int \diff \chi \diff \chi' \frac{W^y(\chi)K(\chi,\chi')W^\kappa(\chi)W^\kappa(\chi')}{d_A(\chi)^6 d_A({\chi'})^6} \int \frac{\diff^2 \vec {\ell'}}{(2\pi)^2} \\
	&\quad\times \abs{\vec{\ell'}}^4\abs{\vec\ell_1 - \vec{\ell'}}^2 \left[\vec{\ell'}\left(\vec\ell_1 - \vec{\ell'}\right)\right] \Pp\left(\frac{\abs{\vec\ell'}}{d_A(\chi)},\chi\right)\Pp\left(\frac{\abs{\vec\ell_1 - \vec{\ell'}}}{d_A(\chi')},\chi'\right)\\
	&\quad\times \left[\cos(2\phi_{\ell_2} - 2\phi_{\vec{\ell'}}) + \cos(2\phi_{\ell_2} - 2\phi_{\vec\ell_1 - \vec{\ell'}})\right] \ .
\end{splitequation}
The dominant contribution is the third order correction \eqref{equ:3rd-order-rs}, by virtue of being of a lower order than the other contributions considered in this work, which are all of fourth order.

\subsection{Redshift distortions}
\label{sec:redshift-distortions}
The comoving line-of-sight distance to a source is usually not an observable quantity. Instead it is derived from the measured redshift, which is affected by the peculiar motions of the source and observer, Sachs-Wolf, and integrated Sachs-Wolf effects.
The second order contribution to the convergence due to a perturbation of the cosmological redshift is
\begin{splitequation}
	\label{equ:kappa-2-z}
	\kappa^{(2)}_z(\chi) = \frac{\diff \kappa^{(1)}(\chi)}{\diff z}\delta z^{(1)} = \frac{\diff \kappa^{(1)}(\chi)}{\diff \chi}\frac{\diff \chi}{\diff z}\delta z^{(1)} \ .
\end{splitequation}
The dependence of the convergence on comoving distance of the source is
\begin{splitequation}
	\frac{\diff \kappa^{(1)}(\chi)}{\diff \chi} = \frac{1}{d_A(\chi)^2}\int_0^{\chi}\diff \chi' \frac{\Phi_{,ii}(\chi')}{d_A(\chi')^2} \ ,
\end{splitequation}
while the redshift perturbation due to peculiar motion of the source, Sachs-Wolf, and integrated Sachs-Wolf effects is given by \cite{Bernardeau2010}
\begin{splitequation}
	\label{equ:z-perturbation}
	\delta z^{(1)} = \frac{1}{a}\left( -2\int_0^{\chi}\diff \chi' \frac{\partial \Phi(\chi')}{\partial \chi'} + \Phi(\chi) - n^i v^{(1)}_i(\chi) \right) \ ,
\end{splitequation}
where the potential at the observer and the peculiar motion of the observer have been set to zero, as they would only contribute at the very large scales. The peculiar motion from first order perturbation theory is \cite{Bernardeau2012}
\begin{splitequation}
	\label{equ:peculiar-velocity}
	v^{(1)}_i(\chi) = - \frac{2a}{3 H_0^2 \Omega_m} \partial_i\left(-\frac{\partial \Phi(\chi')}{\partial \chi'} +\cH(\chi)\Phi(\chi)\right) \ .
\end{splitequation}
From \eqref{equ:z-perturbation} and \eqref{equ:peculiar-velocity} we can already see that only the term corresponding to the peculiar motion would contribute appreciably as it involves an angular derivative. Restricting ourselves to the contribution due to the peculiar motion, the second order convergence can be written as
\begin{splitequation}
	\kappa^{(2)}_z(\chi) = \frac{-  n^i v^{(1)}_i(\chi) }{\cH(\chi)d_A(\chi)^2}\int_0^{\chi}\diff \chi' \frac{\Phi_{,ii}(\chi')}{d_A(\chi')^2} \ .
\end{splitequation}
The photon trajectory $\vec n$ projects the peculiar velocity along the line-of-sight, i.e., the angular derivatives are projected out. Thus all redshift distortions that contribute to \eqref{equ:kappa-2-z} have only two angular derivatives and can be safely neglected on small scales.

\subsection{Vector modes}
\label{sec:vector-modes}
In \cite{Andrianomena2014} it was shown that fourth order contributions from vector modes to lensing observables can be of comparable magnitude as other fourth order contributions considered in this work. It would thus be conceivable that there are large third order contributions involving vector modes. The lowest order cross correlation that includes vector modes is $\braket{\hat y^{(1)}(\vec\ell_1)\hat\kappa^{(2)}_v(\vec\ell_2)}$, where the second order contribution to the convergence is \cite{Andrianomena2014,Thomas2014}
\begin{splitequation}
	\kappa^{(2)}_v(\chi) = \int_0^{\chi} \diff \chi' K(\chi,\chi') n_j V_{,ii}^j(\chi') \ .
\end{splitequation}
The contraction of the line-of-sight direction $\vec n$ with the vector potential $V^i$ is proportional to
\begin{splitequation}
	n_i V^i(\chi) \propto \sin \vartheta\ \e^{\pm \imath \varphi} \ ,
\end{splitequation}
where $\vartheta$ and $\varphi$ denote the spherical coordinates on the sky. This expression is manifestly of odd parity and does not contribute if one correlates it with the even parity field $y$. The lowest order vector contribution to the cross power spectrum has to be quadratic in the vector potential. Since the vector potential is already of second order in the scalar potential $\Phi$, and the lowest vector contribution to $y$ is of third order, there are no fourth order vector contribution to the cross power spectrum.

\section{Discussion}
\begin{figure}
	\includegraphics[width=\textwidth]{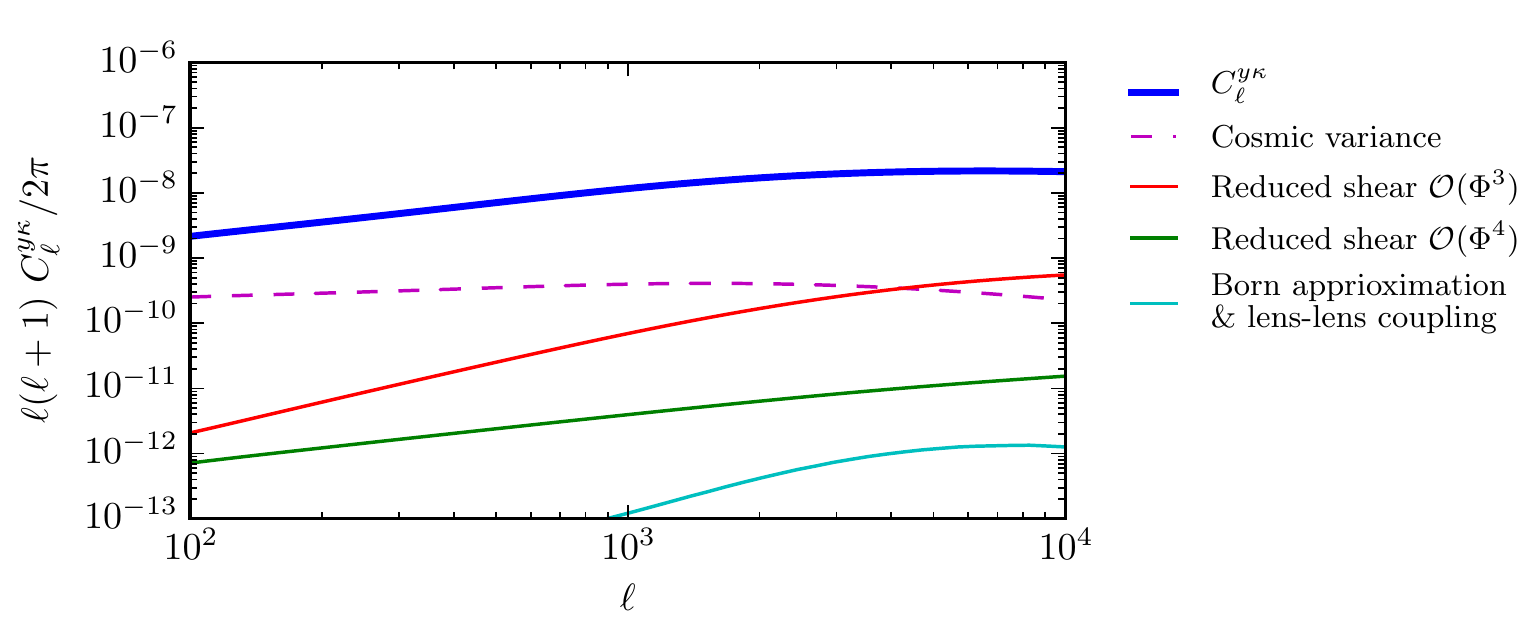}
	\caption{The different contributions to the angular cross power spectrum $C^{y\kappa}_\ell$. The first order result \eqref{equ:y-kappa-1st}(bold blue), third order reduced shear \eqref{equ:3rd-order-rs}(red), fourth order Born and lens-lens terms \eqref{equ:y2-kappa2} and \eqref{equ:y1-kappa3}(cyan), and fourth order reduced shear contributions \eqref{equ:y-kappa-3-RS-A}, \eqref{equ:y1-kappa3-RS-B}, \eqref{equ:y1-kappa3-RS-C}, \eqref{equ:y2-kappa2-RS} (green). \label{fig:corrections}}
\end{figure}

\begin{figure}
	\includegraphics[width=\textwidth]{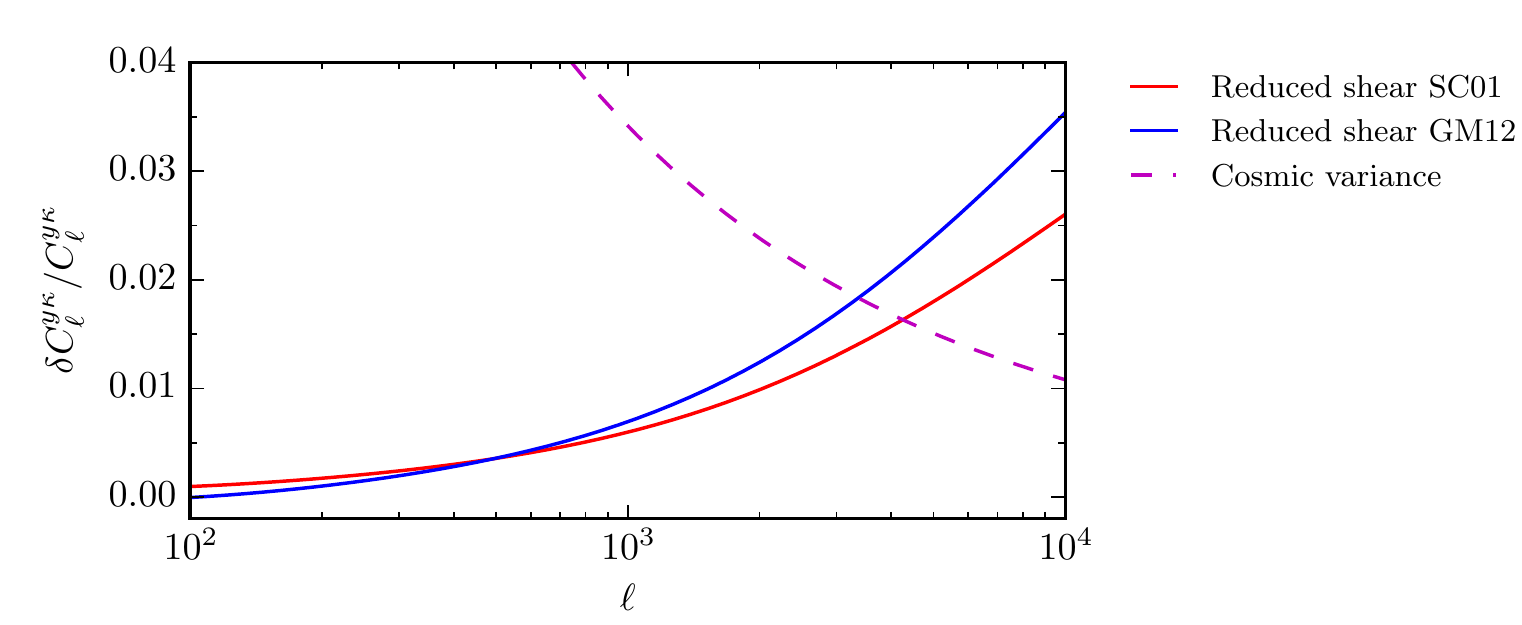}
	\caption{The third order contribution \eqref{equ:3rd-order-rs} to the cross power spectrum computed using the fitting formulae for the bispectrum from \cite{Scoccimarro2001a} and \cite{Gil-Marin2012}. The corrections begin to dominate over cosmic variance above $\ell \sim 4000$. \label{fig:bispec-fit}}
\end{figure}

In figure \ref{fig:corrections} we have plotted the cross power spectrum \eqref{equ:y-kappa-1st} and the various higher order contributions considered in this work. The underlying non-linear matter power spectrum was computed with CAMB\footnote{\url{http://camb.info}}, using the best fit Planck cosmological parameters \cite{PlanckCollaboration2013}. For the source redshift distribution $n(z)$ we use the fitting formula of the redshift distribution of CFHTLenS \cite{VanWaerbeke2013a}. We computed the non-linear bispectrum using the fitting formulae of both \cite{Scoccimarro2001a} and \cite{Gil-Marin2012}. It was found in \cite{Fu2014} that the fitting formula \cite{Gil-Marin2012} slightly overestimates the bispectrum on small scales compared to \cite{Scoccimarro2001a}. For clarity, we only show the reduced shear contribution computed with the fitting formula \cite{Scoccimarro2001a} in figure \ref{fig:corrections}. The relative contributions to the cross power spectrum due to the third order term \eqref{equ:3rd-order-rs} with both fitting formulae is shown in figure \ref{fig:bispec-fit}. 
We find that the third order contribution \eqref{equ:3rd-order-rs} gives the largest correction to the cross power spectrum. At multipoles of $\ell \sim 4000$ it begins to dominate over the cosmic variance $\left((C^{y\kappa}_\ell)^2 + C^{yy}_\ell C^{\kappa\kappa}_\ell \right)/(2\ell + 1)$ \cite{Kamionkowski1997} and reaches multiple percents of the second order result \eqref{equ:y-kappa-1st} at multipoles of $\ell \sim 10^4$. The fourth order contributions are over an order of magnitude lower at small scales. Furthermore, the difference between the two fitting formulae for the bispectrum \cite{Scoccimarro2001a} and \cite{Gil-Marin2012} are at least an order of magnitude larger on small scales than the fourth order corrections. It is thus justified to approximate the fourth order contribution \eqref{equ:y-kappa-3-RS-A-verbose} by its unconnected part, as it is expected to dominate over the connected part at all but the smallest scales \cite{Krause2010}.

\section{Conclusion}
We have calculated all contributions up to fourth order due to weak lensing to the tSZ-lensing cross correlation in the small angle approximation. We have found that only the third order term \ref{equ:3rd-order-rs} due to the reduced shear contributes appreciably. At multipoles of $\ell \sim 3000$ the contribution reaches percent level and raises strongly from there. The effect might thus be observable in future high-resolution surveys, in particular for cluster samples where the tSZ-lensing cross correlation signal will be measured around clusters and stacked.
For upcoming large-area surveys such as LSST\footnote{\url{http://www.lsst.org}} and Euclid\footnote{\url{http://sci.esa.int/euclid/}}, a full-sky treatment will be necessary. As is evident from the large amount of terms in even the second order shear in \cite{Bernardeau2010,Bernardeau2012}, a derivation to the same order as considered in this work will be a formidable task.

Even though the simple bias model employed in this work is compatible with a halo model approach \cite{Ma2014}, a treatment of the corrections considered in this work in the context of the halo model would be of interest. It should be noted that even within the framework of the halo model there is still considerable uncertainty in the modelling of the pressure profile, exemplifying the complications one encounters once baryonic physics are introduced.

Our work can be used to calculate high order lensing corrections to cross correlation signals using a continuous map other than tSZ. For instance one can envision measuring the cross correlation between the Cosmic Infrared Background and lower redshift structures.

\acknowledgments
We thank Alex Hall for helpful comments. TT is supported by the Natural Sciences and Engineering Research Council of Canada (NSERC). LVW is supported by NSERC and the Canadian Institute for Advanced Research (CIfAR).

\appendix
\section{Fourier space identities}
\label{sec:fourier-identities}
Adapting the notation of \cite{Dodelson2005a}, we define the 2d Fourier transform on the plane perpendicular to the line-of-sight as
\begin{splitequation}
	\label{equ:def-fourier-transform}
	\Phi(d_A(\chi)\vec\theta, \chi) &=  \int \frac{\diff^2 \vec {\ell'}}{(2\pi)^2 } \hat\phi(\vec {\ell'}, \chi) \e^{\imath \vec {\ell'}\vec\theta} \ ,
\end{splitequation}
with the angular transform of the field $\Phi$ given by
\begin{equation}
	\label{equ:def-angular-fourier-transfrom}
	\hat \phi(\vec \ell,\chi) = \int \frac{\diff k_3}{2\pi} \frac{1}{d_A(\chi)^2}\hat\Phi\left(\frac{\vec\ell}{d_A(\chi)},k_3\right)\e^{\imath k_3 \chi} \ .
\end{equation}
The higher order expressions for $y$ and $\kappa$ involve products of the potential $\Phi$. In Fourier space, these products become convolutions. For two fields F and G we have
\begin{splitequation}
\label{equ:convolution-def}
	\widehat{[FG]}(\vec\ell) = \left[\hat F * \hat G\right](\vec\ell) = \int \frac{\diff^2 \vec{\ell'}}{(2\pi)^2}\hat F(\vec{\ell'})\hat G(\vec\ell - \vec{\ell'}) \ .
\end{splitequation}
This generalizes straightforwardly to the case of three fields F,G, and K as
\begin{splitequation}
\label{equ:3-convolution-def}
	\widehat{[FGK]}(\vec\ell) = \left[\hat F * \hat G * \hat K\right](\vec\ell) = \int \frac{\diff^2 \vec{\ell'}\diff^2 \vec{\ell''}}{(2\pi)^4}\hat F(\vec{\ell'})\hat G(\vec{\ell''})\hat K(\vec\ell - \vec{\ell'} - \vec{\ell''}) \ .
\end{splitequation}
The two point correlation function of the fields $\hat\phi(\vec\ell, \chi)$ is directly related to the power spectrum of the potential $\Phi$. Assuming homogeneity, isotropy, and using the Limber approximation \cite{Limber1953, Kaiser1992}, i.e., assuming that $\abs{\vec\ell} \gg k_3$, thus justifying neglecting the longitudinal modes, the two point function can be written as
\begin{splitequation}
\label{equ:powerspectrum-def}
	\braket{\hat\phi(\vec\ell_1,\chi_1)\hat\phi(\vec\ell_2, \chi_2)} &= (2\pi)^2 \ddelta(\chi_1 - \chi_2) \ddelta(\vec\ell_1 + \vec\ell_2) C^{\phi\phi}_\ell \\
	&= (2\pi)^2 \ddelta(\chi_1 - \chi_2)\frac{\ddelta^2\left(\vec\ell_1+\vec\ell_2\right)}{d_A(\chi_1)^2 } \Pp\left(\frac{\abs{\vec\ell_1}}{d_A(\chi_1)}, \chi_1\right) \ .
\end{splitequation}
Similarly, the three-point function of $\hat\phi(\vec\ell, \chi)$ is related to the bispectrum $\Bp(\vec k_1, \vec k_2, \vec k_3, \chi)$ by
\begin{multline}
\label{equ:bispectrum-def}
	\braket{\hat\phi(\vec\ell_1,\chi_1)\hat\phi(\vec\ell_2, \chi_2)\hat\phi(\vec\ell_3, \chi_3)} = (2\pi)^2 \ddelta(\chi_1 - \chi_2)\ddelta(\chi_1 - \chi_3)\frac{\ddelta^2\left(\vec\ell_1+\vec\ell_2 + \vec\ell_3\right)}{d_A(\chi_1)^4 } \\
	\times \Bp\left(\frac{\abs{\vec\ell_1}}{d_A(\chi_1)},\frac{\abs{\vec\ell_2}}{d_A(\chi_1)},\frac{\abs{\vec\ell_3}}{d_A(\chi_1)}, \chi_1\right) \ .
\end{multline}
The four-point function can be expressed in terms of the trispectrum as
\begin{splitequation}
\label{equ:trispectrum-def}
	\braket{\hat\phi(\vec\ell_1,\chi_1)\hat\phi(\vec\ell_2, \chi_2)\hat\phi(\vec\ell_3, \chi_3)\hat\phi(\vec\ell_4, \chi_4)} &= (2\pi)^2 \ddelta(\chi_1 - \chi_2)\ddelta(\chi_1 - \chi_3)\ddelta(\chi_1 - \chi_4) \\
				&\quad\times \frac{\ddelta^2\left(\vec\ell_1+\vec\ell_2 + \vec\ell_3 + \vec\ell_4\right)}{d_A(\chi_1)^6 } \\
				&\quad\times \cT_\Phi\left(\frac{\abs{\vec\ell_1}}{d_A(\chi_1)},\frac{\abs{\vec\ell_2}}{d_A(\chi_1)},\frac{\abs{\vec\ell_3}}{d_A(\chi_1)}, \frac{\abs{\vec\ell_4}}{d_A(\chi_1)}, \chi_1\right) \ .
\end{splitequation}	
Using the definition of the Fourier transform \eqref{equ:def-fourier-transform}, partial derivatives with respect to comoving transverse coordinates can be written as
\begin{splitequation}
	\label{equ:fourier-deriv}
	\Phi_{,i_1 \dots i_N}(d_A(\chi)\vec\theta, \chi) = \int \frac{\diff^2 \vec {\ell'}}{(2\pi)^2} \frac{\imath^N}{d_A(\chi)^N} {\ell'}_{i_1}\dots {\ell'}_{i_N}\hat\phi(\vec {\ell'}, \chi) \e^{\imath \vec {\ell'}\vec\theta} \ .
\end{splitequation}

\section{Convergence - shear relation}
\label{sec:gamma-kappa-rel}
In this appendix we show that the relation \eqref{equ:kappa-gamma-rel} holds even beyond first order, justifying the use of the convergence as the fundamental quantity instead of the shear. The second order expressions of the convergence and shear due to Born approximation and lens-lens couplings in Fourier space are
\begin{splitequation}
	\hat\kappa^{(2)}_{std}(\vec\ell, \chi) &= -2\int_0^{\chi}\diff \chi' \int_0^{\chi'}\diff \chi'' \frac{K(\chi,\chi')K(\chi',\chi'')}{d_A(\chi')^2d_A(\chi'')^2} \\
	&\quad\times  \int \frac{\diff^2\vec {\ell'}}{(2\pi)^2} \abs{\vec {\ell'}}\abs{\vec\ell}\cos(\phi_{\ell} - \phi_{\ell'})\left[\vec {\ell'}(\vec\ell - \vec {\ell'}) \right] \hat\phi(\vec {\ell'}, \chi')\hat\phi(\vec\ell - \vec {\ell'}, \chi'')
\end{splitequation}
and
\begin{splitequation}
	(\hat\gamma^{(2)}_{std})_I(\vec\ell, \chi) &= -2\int_0^{\chi}\diff \chi' \int_0^{\chi'}\diff \chi'' \frac{K(\chi,\chi')K(\chi',\chi'')}{d_A(\chi')^2d_A(\chi'')^2} \\
	&\quad\times  \int \frac{\diff^2\vec {\ell'}}{(2\pi)^2} \abs{\vec {\ell'}}\abs{\vec\ell}U_I(\vec\ell, \vec{\ell'})\left[\vec {\ell'}(\vec\ell - \vec {\ell'}) \right] \hat\phi(\vec {\ell'}, \chi')\hat\phi(\vec\ell - \vec {\ell'}, \chi'') \ ,
\end{splitequation}
where the couplings $U_I$ are given by
\begin{splitequation}
	U^1(\vec\ell, \vec {\ell'}) = \cos(\phi_\ell + \phi_{\ell'}) \ ,\quad U^2(\vec\ell, \vec {\ell'}) =  \sin(\phi_\ell + \phi_{\ell'}) \ .
\end{splitequation}
Using the identity
\begin{splitequation}
	\label{equ:T-Q-identity}
	T^I(\vec\ell)U_I(\vec\ell, \vec {\ell'}) = \cos(\phi_\ell - \phi_{\ell'}) \ ,
\end{splitequation}
where $T^I$ is given in \eqref{equ:kappa-gamma-rel}, we have thus shown that $\hat\kappa^{(2)}_{std} = T^I(\hat\gamma^{(2)}_{std})_I$.

Generally, the relation does not hold anymore at third order. As we are only concerned with correlation functions in this work, it is sufficient to show that the relation holds within correlation functions, i.e., $\braket{\hat y^{(1)}\hat\kappa^{(3)}_{std}}=\braket{\hat y^{(1)}T^I(\hat\gamma^{(3)}_{std})_I}$. To do so we first note that the third term in the third order expression for the convergence \eqref{equ:kappa-3rd-order} does not contribute to the correlation function under the Limber approximation. This also applies to the third order shear, since the line-of-sight integrals are the same for both the convergence and the shear. The mode coupling terms of the angular cross power spectra are
\begin{splitequation}
	 \braket{\hat y^{(1)}(\vec\ell_1)\hat\kappa^{(3)}_{std}(\vec\ell_2)}\propto \int \frac{\diff^2\vec {\ell'}}{(2\pi)^2}\left(\abs{\vec {\ell_2}}^2 + 2\vec {\ell_2}\vec {\ell'}\right) \abs{\vec\ell_1}^2\left[\vec {\ell_2}\vec {\ell'}\right]^2  
\end{splitequation} 
and
\begin{splitequation}
	 \braket{\hat y^{(1)}(\vec\ell_1)T^I(\vec\ell_2)(\hat\gamma^{(3)}_{std})_I(\vec\ell_2)}\propto T^I(\vec\ell_2)\int \frac{\diff^2\vec {\ell'}}{(2\pi)^2}\left(\abs{\vec {\ell_2}}^2 T_I(\vec\ell_2) - 2\abs{\vec {\ell_2}}\abs{\vec {\ell'}}U_I(\vec\ell_2,\vec{\ell'})\right)  \abs{\vec\ell_1}^2\left[\vec {\ell_2}\vec {\ell'}\right]^2 \ .
\end{splitequation} 
Applying the two identities $T^I(\vec\ell)T_I(\vec {\ell'}) = \cos(2\phi_{\ell'} - 2\phi_\ell)$ and \eqref{equ:T-Q-identity} we see that the two above expressions are equal. We have thus proven that it is justified to use the convergence in cross power spectra instead of terms of the from $T^I(\vec\ell)\hat\gamma_I(\vec\ell)$, up to third order. To second order, the relation $\hat\kappa = T^I\hat\gamma_I$ even holds exactly.

\section{Induced rotation}
\label{sec:rotation}
Let $S(\vec\theta)$ be the surface brightness distribution of an extended source. The first and second moment of the brightness distributions are then defined as \cite{Schneider1995}
\begin{subequations}
	\begin{align}
		\theta^0_i &= \frac{\int\diff^2\vec\theta\ \theta_i S(\vec\theta)}{\int\diff^2\vec\theta\ S(\vec\theta)} \\ 
		Q_{ij} &= \frac{\int\diff^2\vec\theta\ (\theta_i-\theta^0_i)(\theta_j-\theta^0_j) S(\vec\theta)}{\int\diff^2\vec\theta\ S(\vec\theta)} \label{equ:Q-def}\ .
	\end{align} 
\end{subequations}
Following \cite{Seitz1997}, we introduce the complex ellipticity parameter
\begin{splitequation}
	\label{equ:epsilon-def}
	\epsilon = \frac{Q_{11}-Q_{22}+2\imath Q_{12}}{Q_{11}+Q_{22}+2\sqrt{Q_{11}Q_{22}-Q_{12}^2}} \ .
\end{splitequation}
For an elliptical source with semi major and minor axis $a$ and $b$, rotated by an angle $\alpha$ with respect to a fixed coordinate system, the ellipticity parameter \eqref{equ:epsilon-def} is given by
\begin{splitequation}
	\epsilon = \frac{a-b}{a+b}\e^{2\imath\alpha} \ .
\end{splitequation}
The Jacobi map \eqref{equ:distortion-matrix} relates an infinitesimal distance on the source plane to an infinitesimal distance on the image plane by $\diff \vec\theta^S=\cA(\vec\theta^S)\ \diff\vec\theta^O$. Assuming the source is sufficiently small such that the Jacobi map does not vary over the extend of the source, the second brightness moment \eqref{equ:Q-def} of the source $Q^S$ can be approximately related to the observed second brightness moment $Q^O$ by
\begin{splitequation}
	Q^S=\cA\ Q^O \cA^T \ .
\end{splitequation}
We generalize previous work by allowing $\cA(\vec\theta)$ to have an anti-symmetric part. This anti-symmetric contribution $\omega$ can be thought of as a rotation induced by lens-lens coupling. Given an elliptical source, the observed ellipticity can be written as
\begin{splitequation}
	\label{equ:epsilon-I}
	\epsilon^O = \frac{g+{\epsilon^S}'}{1+\conj{g}{\epsilon^S}'} \ ,
\end{splitequation}
where the generalized reduced shear $g$ and rotated source ellipticity ${\epsilon^S}'$ are given by
\begin{splitequation}
	\label{equ:general-g}
	g=\frac{\gamma_1+\imath\gamma_2}{1-\kappa+\imath\omega}\ , \quad {\epsilon^S}' = \epsilon^S\ \e^{-2\imath\vartheta}\ , \quad \tan\vartheta = \frac{\omega}{1-\kappa} \ .
\end{splitequation}
The reduced shear now includes a contribution from the anti-symmetric term $\omega$ of the general Jacobi map. Furthermore, the source ellipticity is rotated by an angle $\vartheta$. However, assuming the sources are distributed isotropically, this rotation is not observable. In particular, the ensemble average $\braket{\epsilon^O}=g$, i.e. the observed ellipticity remains an unbiased estimator of the reduced shear despite the rotation $\omega$.
In the limit of a symmetric Jacobi map $\omega\rightarrow 0$ one recovers eq. (3.2) of \cite{Seitz1997}.

Finally, we express the generalized reduced shear in vector notation to facilitate the use in section \ref{sec:reduced-shear}. The two components are
\begin{splitequation}
	g_1=\Re{(g)}=\frac{\gamma_1(1-\kappa) + \gamma_2\omega}{(1-\kappa)^2+\omega^2}\ , \quad
	g_2=\Im{(g)}=\frac{\gamma_2(1-\kappa) - \gamma_1\omega}{(1-\kappa)^2+\omega^2} \ .
\end{splitequation}
Because $\omega$ is necessarily of at least second order, the generalized reduced shear to third order can be written as
\begin{splitequation}
	\label{equ:general-g-approx}
	g_I=\frac{\gamma_I}{1-\kappa}+R(\omega)_{IJ}\gamma_J + \cO(\Phi^4) \ ,
\end{splitequation}
where the matrix $R(\omega)$ is defined as
\begin{equation}
	\label{equ:R-omega}
	R(\omega)_{IJ} = 
	\begin{pmatrix}
		0 & \omega \\
		-\omega & 0
	\end{pmatrix} \ .
\end{equation}
This can be understood as an infinitesimal rotation of the shear by an angle $\omega$.

\bibliographystyle{jhep}
\bibliography{y-kappa.bib}

\end{document}